\documentclass[conference]{IEEEtran}
\IEEEoverridecommandlockouts
\usepackage{cite}
\usepackage{amsmath,amssymb,amsfonts}
\usepackage{algorithmic}
\usepackage{graphicx}
\usepackage{textcomp}
\usepackage{xcolor}
\usepackage{booktabs} 
\usepackage[linesnumbered,ruled]{algorithm2e}
\usepackage{multirow}
\usepackage{caption}
\usepackage{subcaption}
\def\BibTeX{{\rm B\kern-.05em{\sc i\kern-.025em b}\kern-.08em
    T\kern-.1667em\lower.7ex\hbox{E}\kern-.125emX}}
\begin{document}

\title{Top-N-Rank: A Scalable List-wise Ranking Method for Recommender Systems\\
\thanks{Part of this work was conducted when the first author was at the South China University of Technology.}
\thanks{\IEEEauthorrefmark{1}Corresponding author (jlhu@scut.edu.cn).}
}

\author{\IEEEauthorblockN{Junjie Liang\IEEEauthorrefmark{2} ,
Jinlong Hu\IEEEauthorrefmark{3}\IEEEauthorrefmark{1}, Shoubin Dong\IEEEauthorrefmark{3} and
Vasant Honavar\IEEEauthorrefmark{2}}
\IEEEauthorblockA{
\IEEEauthorrefmark{2}Artificial Intelligence Research Laboratory, Center for Big Data Analytics and Discovery Informatics\\College of Information Sciences and Technology, 
Pennsylvania State University, University Park, PA 16802, United States\\
  Email: \{jul672, vhonavar\}@ist.psu.edu\\}
  \IEEEauthorblockA{
\IEEEauthorrefmark{3}Guangdong Key Laboratory of Communication \& Computer Network, School of Computer Science and Engineering\\
South China University of Technology, Guangzhou 510006, China\\
  Email: \{jlhu, sbdong\}@scut.edu.cn
}
}

\maketitle

\begin{abstract}
   We propose Top-N-Rank, a novel family of list-wise Learning-to-Rank models for reliably recommending the N top-ranked items. The proposed models optimize  a  variant of the widely used discounted cumulative gain (DCG) objective function which differs from DCG in two important aspects: (i) It limits the evaluation of DCG only on the top N items in the ranked lists, thereby eliminating the impact of low-ranked items on the learned ranking function;  and  (ii) it incorporates weights that allow the model to leverage multiple types of implicit feedback with differing levels of reliability or trustworthiness. Because the resulting objective function is non-smooth and hence challenging to optimize, we consider  two smooth approximations of the objective function, using the traditional sigmoid function and the rectified linear unit (ReLU). We propose a family of learning-to-rank algorithms (Top-N-Rank) that work with any smooth objective function. Then, a more efficient variant, Top-N-Rank.ReLU, is introduced, which effectively exploits the properties of ReLU function to reduce the computational complexity of Top-N-Rank from quadratic to linear in the average number of items rated by users. The results of our experiments using two widely used benchmarks, namely, the MovieLens data set and the Amazon Video Games data set demonstrate that: (i) The ``top-N truncation'' of the objective function substantially improves the ranking quality of the top N recommendations; (ii) using the ReLU for smoothing the objective function yields significant improvement in both ranking quality as well as runtime as compared to using the sigmoid; and (iii) Top-N-Rank.ReLU substantially outperforms the well-performing list-wise ranking methods in terms of ranking quality.  
\end{abstract}

\begin{IEEEkeywords}
list-wise ranking; top n recommendation; learning to rank; latent factor model; rectifier function;

\end{IEEEkeywords}

%
\IEEEpeerreviewmaketitle

\section{Introduction}
Collaborative filtering (CF) is one of the most widely used methods in recommender systems. CF systems recommend similar items to users who share similar traits or similar tastes \cite{gunawardana_survey_2009}.  Learning to Rank (LTR) methods, which directly learn to accurately rank items based on the user's ratings, rankings, or implicit feedback over a set of items, are widely used to learn the perfect rankings for top-n recommendation scenarios \cite{liu_learning_2009,shi_climf:_2012}. 

\subsection{Related Work}
Learning to Rank (LTR) methods can be categorized into point-wise, pair-wise and list-wise methods. Point-wise methods learn ranking models from the scores assigned by users to individual items (see e.g., \cite{he_neural_2017}). Pair-wise methods (e.g., BPR \cite{rendle_bpr:_2009}), learn binary classifiers that compare ordered pairs of items to decide whether the first item is preferred to the second. The applicability of such methods is limited by the high computational cost of pair-wise comparisons of user-rated items in generating the training samples for the binary classifiers \cite{huang2015listwise}. List-wise methods leverage the entire list of items consumed by the users to optimize a list-wise ranking loss function or the probability of permutations that map items to ranks \cite{tie-yan_learning_2014,xia_listwise_2008}. Typically, such methods optimize a smooth approximation of a loss function that measures the distance  between the reference lists of ranked items in the training data and the ranked list of items produced by the ranking model. For example, CLiMF \cite{shi_climf:_2012}, which optimizes a smooth lower bound of mean-reciprocal rank (MRR), aims at ranking a small set of most-preferred items at the top of the list; TFMAP \cite{shi_tfmap:_2012} optimizes the mean average precision (MAP) of top-ranked items for each user in a given context. Other methods optimize discounted cumulative gain (DCG) or normalized DCG (NDCG) can be found in  \cite{lim_top-n_2015,steck_gaussian_2015}. Examples of list-wise methods that optimize the probability of permutations that map items to ranks include: ListPMF, which represents each user as a probability distribution of the permutations over rated items based on the Plackett-Luce model \cite{liu_list-wise_2014}; ListRank \cite{shi_list-wise_2010}, which aims to identify a ranking permutation that minimizes the cross-entropy between the distribution of the observed ranking of items based on user ratings and the predicted rankings with respect to the top-ranked item; or methods that optimize the log-posterior over the predicted preference order with the observed preference orders \cite{liu_list-wise_2014}; and methods that leverage deep neural nets (e.g., \cite{he_neural_2017}) to learn the non-linear interaction between user-item pairs (See \cite{zhang2017deep} for a survey of such methods). 

Existing LTR approaches suffer from several limitations. Although, in practical applications, only the top (say N) items in the ranked list are of interest, and the lower-ranked ratings in the list are less reliable, most existing LTR methods are optimized on the ranks of the entire lists, which, could potentially reduce the ranking quality of the top-ranked items. Furthermore, the computational complexity of straightforward approaches to optimizing ranking measures (e.g., DCG \cite{ifada_-rank:_2015}, MRR \cite{shi_climf:_2012}, AUC \cite{steck_gaussian_2015} or MAP \cite{shi_tfmap:_2012}), scale quadratically with $\tilde{m}$ (the average number of observed items across all users), which renders such methods impractical in large-scale real-world settings.

\subsection{Overview and Contributions}
To address the limitations of existing LTR systems, we propose Top-N-Rank, a novel latent factor based list-wise ranking model for top-N recommendation problem which directly optimizes a novel weighted ``top-heavy'' truncated variant of the DCG ranking measure, namely,  wDCG@N. Since in many situations, the users only refer to the top-ranked items in the list, the higher positions often have more impact on the ranking score than the lower ones. Our proposed measure, wDCG@N,  differs from the conventional DCG in two important aspects: (i) It considers only on the top N items in the ranked lists, thereby eliminating the impact of low-ranked items; and  (ii) it incorporates weights that allow the model to learn from multiple kinds of implicit feedback.

Because wDCG@N is non-smooth, we introduce the rectified linear unit (ReLU) as a smoothing function, which is more suited to top N ranking problems than the traditional sigmoid function. ReLU not only eliminates the contribution of the low-ranked items on our loss function, but also allows us to obtain  a significantly faster variant of the wDCG@N-based LTR approach (Top-N-Rank.ReLU), yielding a substantial reduction in computational complexity from $O(kn'\tilde{m}^2)$ to $O(kn'\tilde{m})$, where $k$ denotes the dimension of latent factors, $n'$ denotes the batch size of users for the stochastic gradient descent algorithm and $\tilde{m}$ is the average number of (observed) items, making Top-N-Rank.ReLU scalable to large-scale real-world settings.

The main contributions of this paper can be summarized as follows:
\begin{enumerate}
  \item We have introduced a novel list-wise ranking model for top-N recommendation, which directly optimizes a weighted top-heavy truncated ranking objective function, wDCG@N. Our model improves the quality of the top-N item lists by mitigating the impact of the lower-ranked items, and is capable of handling multiple types of implicit feedback (if available).
  \item We have introduced the rectified linear unit (ReLU) to smooth our objective function. We have demonstrated that ReLU could (1) eliminate the impact of the lower-ranked items and (2) substantially speed up the calculations by careful algorithm design.
  \item We have proposed a fast learning algorithm (Top-N-Rank) for generic smoothing functions, and a substantially more efficient variant (Top-N-Rank.ReLU) for the ReLU smoothing function which reduce the computational complexity from $O(kn'\tilde{m}^2)$ to $O(kn'\tilde{m})$.
\end{enumerate}

We compared the performance of Top-N-Rank and Top-N-Rank.ReLU with several state-of-the-art list-wise LTR methods \cite{lim_top-n_2015,shi_climf:_2012,shi_xclimf:_2013,shi_list-wise_2010,liu_list-wise_2014} using the MovieLens (20M) data set \cite{harper_movielens_2016} and the Amazon video games data set \cite{mcauley_addressing_2016}. All experiments were performed on Apache Spark cluster \cite{shoro_big_2015} and the raw data are stored on Hadoop Distribute File System (HDFS).

\section{Preliminaries}
Let $\mathcal{U}=\left\{u_1,u_2,\ldots,u_n\right\}$ be the set of $n$ users, $\mathcal{I}=\left\{i_1,i_2,\ldots,i_m\right\}$ be the set of $m$ items and $\mathcal{P}=\left\{p_1,p_2,\ldots,p_t\right\}$ be the set of $t$ types of implicit feedback. The interactions of users with items and the associated implicit feedback are represented by $\mathcal{X}=\mathcal{U}\times\mathcal{I}\times\mathcal{P}$, where the entry $\left(u,i,p_{ui}\right)\in\mathcal{X}$ denotes the interaction of user $u$ with item $i$ and the associated implicit feedback $p_{ui}$. We further denote by $\mathcal{I}_u^+$ the subset of items actually observed by or presented to $u$. For each $i\in\mathcal{I}_u^+$, we denote the rating of $i$ by $f_{ui}$ and the position of $i$ based on $u$'s rank ordering of items by $R_{ui}^+$. We reserve the indexing letters $u$ to indicate arbitrary user in $\mathcal{U}$ and $i$ to represent arbitrary item in $\mathcal{I}$.

\subsection{Latent Factor Model}
Latent factor models (LFMs) are state-of-the-art in terms of both the quality of recommendations as well as scalability \cite{aggarwal_model-based_2016}. LFMs represent both users and items using low-dimensional vectors of latent factors. Let $\theta$ be the set of latent factors such that $\theta^{user}$ is an $n\times k$ matrix with the $u$-th row $\theta_u^{user}$ denoting the latent factors of $u$ and $\theta^{item}$ is an $m\times k$ matrix with the $i$-th row $\theta_i^{item}$ denoting the latent factors of $i$. The rank $k$, of the latent factor matrices is much smaller than $n$ or $m$. The rating for $u$ to $i$ is predicted by the dot product of $\theta_u^{user}$ and $\theta_i^{item}$.

\subsection{Discounted Cumulative Gain}
The discounted Cumulative Gain (DCG) \cite{jarvelin2002cumulated} is a widely used measure of quality of recommendations, which measures the degree to which higher ranked items are placed ahead of the lower ranked ones, with the contribution of lower ranked items discounted by a logarithmic factor.
Let $y_{ui}$ be a binary indicator to represent whether $i$ is relevant to $u$, then DCG of $u$ is computed by:

\begin{equation}
  \label{eq2}
  \mathrm{DCG}_u=\sum_{i\in \mathcal{I}}\frac{2^{y_{ui}}-1}{\log{\left(R_{ui}+2\right)}}
\end{equation}

Notice that the ranked position (start from zero) of item $i$ can be computed by:

\begin{equation}
  \label{eq3}
  R_{ui}=\sum_{j\in \mathcal{I}}\mathrm{1}(f_{ui} < f_{uj})
\end{equation}

where $\mathrm{1}(x)$ is an indicator function with $\mathrm{1}(x)=1$ if $x$ is true and otherwise $\mathrm{1}(x)=0$. Given our emphasis on getting the top rated items ranked correctly in the list of recommended items, DCG appears to be good criterion to optimize on. However, as evident from (\ref{eq2}), DCG suffers from two important limitations: (i) Although DCG de-emphasizes the contribution of the lower ranked items, it does not eliminate the collective effect of a large number of lower ranked items, even if the ranking of such lower ranked items are less reliable. If the goal is to optimize the ranking of the N top rated items, it makes sense to tailor objective function to focus explicitly on the ranking of the N top-rated items and ignore the rest. (ii) Because DCG assigns equal weights to all implicit user feedback, it fails to account for differences in their trustworthiness.

\section{Top-N-Rank}
We proceed to introduce wDCG@N, a variant of DCG that overcomes its drawbacks. We then describe two smoothing functions (sigmoid and rectified linear unit (ReLU)) that can convert wDCG@N to a smoothed function that is amenable to being optimized using the standard optimization techniques. Finally, we show how to use the ReLU approximation of wDCG@N to obtain a scalable LTR algorithm. 

\subsection{Top-N-Rank Training Objective}
To address the limitations of DCG, we introduce wDCG@N, which is defined as follows:

\begin{equation} \label{eq4}
  \mathrm{wDCG}_u\mathrm{@N}=\sum_{i\in \mathcal{I}}{\mathrm{1}\left(R_{ui}<N\right)\cdot \frac{w_{p_{ui}}\left(2^{y_{ui}}-1\right)}{\log{\left(R_{ui}+2\right)}}}
\end{equation}

The first term in (\ref{eq4}), $\mathrm{1}\left(R_{ui}<N\right)$ is an indicator function that selects only the N top-rated items and ignores the rest. The coefficient $w_{p_{ui}}$ in the second term denotes the weight of the implicit feedback $p_{ui}$, which can model the reliability or importance of the feedback. The choice of $w_{p_{ui}}$ is application and data-dependent. For example, one can set $w_{p_{ui}}$ to the number of items rated by (or presented to) the user \cite{koren2008factorization} or the conversion rate (the proportion between buyers and users who conducted the implicit feedback). The resulting ranking objective can be formulated as:

\begin{equation}
  \label{eq5}
  L(\theta)=\max_{\theta}{\sum_{u\in\mathcal{U}}{\mathrm{wDCG}_u\mathrm{@N}}-\lambda\left|\left|\theta\right|\right|_2^2}
\end{equation}
where $\left|\left|\cdot \right|\right|_2^2$ denotes the $L^2$-norm and $\lambda$ is the regularization coefficient that controls over-fitting.

\subsection{Smooth Approximations of Top-N-Rank Training Objective}
\label{subsec:smooth_functions}
A non-smooth training objective such as the one in (\ref{eq5}) is challenging to optimize. Hence, we replace the non-smooth training objective in (\ref{eq5}) by its smooth approximation. Specifically, we  approximate the indicator function in (\ref{eq3}) by a smooth function $h$ such that $\mathrm{1}\left(f_{ui}<f_{uj}\right)\approx h\left(\Delta_{uji}\right)$ with $\Delta_{uji}=f_{uj}-f_{ui}$. In what follows, we will consider two different smooth functions that accomplish this goal.

\textbf{Sigmoid function.}~The sigmoid function is widely used in existing list-wise LTR-based recommendation models (e.g., \cite{shi_tfmap:_2012,shi_climf:_2012}) for its appealing performance in practice. Instead of adopting the sigmoid function directly, we introduce a scaling constant $C~(C\geq1)$ to provides more accurate estimation, such that the indicator function is approximated by $g\left(C\Delta_{uji}\right)$ where $g\left(x\right)=1/{\left(1+\exp{\left(-x\right)}\right)}$.

\textbf{Rectifier function.}~The rectified linear units (ReLU) \cite{lecun_deep_2015}, is a nonlinear smooth function with several properties that make it attractive in our setting. The one-sided nature of ReLU ($relu\left(x\right)=\max{\left\{0,x\right\}}$) eliminates the contribution of  of the lower-rated items to the objective function. Second, ReLU is computationally simper: only comparison and addition operations are required. Third, the form of ReLU permits an efficient algorithm (see  Algorithm \ref{alg2}) with computational complexity that is linear in the average number of (observed) items across all users (see section~\ref{sec:Enhance}). When ReLU is used, we have $\mathrm{1}\left(f_{ui}<f_{uj}\right)\approx relu\left(\Delta_{uji}\right)$

\subsubsection{Parameterization of the Smooth Functions}
\label{sec:smooth_func}
Recall that the ``top-N term'', $\mathrm{1}\left(R_{ui}<N\right)$, was introduced to indicate whether item $i$ ranks among the top N item list. However, a poor choice of the hyper-parameters in the smooth function could lead to gross under-estimation or over-estimation of $R_{ui}$ and thus negate the utility of the ``top-N term''.

Here we examine how to choose the parameters of the sigmoid and the ReLU functions so that they behave as intended. In the case of the  sigmoid function, we see that a choice of $C$ matters, with proper values of $C$ (e.g., $C=7$) yielding the desired behavior. In the case of ReLU,  we can ensure the desired behavior by controlling the initial distribution of latent factors $\theta$. Suppose that $\theta\sim U\left(0,b\right)$, where $b$ is the width of the uniform distribution. Then according to the Central Limit Theorem, for arbitrary $u,i$, $f_{ui}$ approximately follows a Gaussian distribution, i.e., $f_{ui}\sim N\left(\mu,\sigma^2\right)$ with $\mu=\mathbb{E}\left[\sum_{t}{\theta_{ut}^{user}\cdot \theta_{it}^{item}}\right]=kb^2/4$ and $\sigma^2=\mathbb{E}\left[\sum_{t}{\theta_{ut}^{user}\cdot \theta_{it}^{item}}\right]^2-\mu^2=7kb^4/144$. In order to ensure that $\left|f_{ui}-\mu\right|\le1$,  making use of  the fact that $P\left(\left|f_{ui}-\mu\right|\le3\sigma\right)\approx1$, we find $3\sigma=1$, and hence $b=2/ \sqrt[4]{7k}$, which provides the basic setting for all of the Top-N models using the ReLU as the smoothing function.

\subsection{Fast LTR Algorithms}
\subsubsection{Fast LTR Algorithm for Generic Smooth Function}
To optimize the objective function reported in (\ref{eq5}), we need to compute the predicted score of each item and then perform the pair-wise comparison to determine their positions in the rank-ordered list. Because in most cases, the number of items $m$ far outnumbers the dimension of the latent factors $k$, the complexity of a single pass is $O\left(knm^2\right)$. One common practice is to exploit the sparsity of $\mathcal{X}$ by considering only the predicted scores of the observed items, yielding a smooth objective function such as:

\begin{equation}
  \label{eq8}
  \begin{aligned}
  L^+(\theta) &= \min_\theta{-\sum_{u\in \mathcal{U}}{\sum_{i\in\mathcal{I}_u^+}{h(N-R_{ui}^+)}}} \\
  & \cdot \frac{w_{p_{ui}}(2^{y_{ui}}-1)}{\log(R_{ui}^+ + 2)}+\lambda||\theta||_2^2
  \end{aligned}
\end{equation}

The gradient of $L^+(\theta)$ w.r.t. $\theta$ is given by (\ref{eq9}).

\begin{equation}
  \label{eq9}
  \begin{aligned}
    \frac{\partial L^+}{\partial \theta}= & \sum_{u \in \mathcal{U}}{\sum_{i\in \mathcal{I}_u^+}{\mathrm{wDCG}_u^+\cdot (\sum_{j\in \mathcal{I}_u^+}{h^\prime(\Delta_{uji})\frac{\partial\Delta_{uji}}{\partial\theta}})}} \\
    & \cdot (h^\prime(N-R_{ui}^+) + \frac{h(N-R_{ui}^+)}{(R_{ui}^+ + 2)\log(R_{ui}^+ + 2)})+ 2\lambda \theta
  \end{aligned}
\end{equation}

\begin{algorithm}
    \SetKwInOut{Input}{Input}
    \SetKwInOut{Output}{Output}
    \Input{User-item feedback $\mathcal{X}\subseteq\mathcal{U}\times\mathcal{I}\times\mathcal{P}$, the truncate coefficient $N$, smooth function $h$, dimension of latent factors $k$, learning rate $\alpha$, regularization coefficient $\lambda$, batch size $n'$}
    \Output{The learned latent factors $\theta$}
	Initialize $\theta^{\left(0\right)}$; set $t=0$ \\
	\While{not converged}{
		$U^{(t)}$ = draw $n'$ samples randomly from $\mathcal{U}$ \\
		\For{$u\in U^{(t)}$}{
		  Update ${\theta_u^{user}}^{\left(t+1\right)}={\theta_u^{user}}^{\left(t\right)}-\alpha\left(\frac{\partial L^+}{\partial{\theta_u^{user}}}\right)^{(t)}$ based on (\ref{eq9}) \\
		  \For{$i\in\mathcal{I}_u^+ $}{
		    Update ${\theta_i^{item}}^{\left(t+1\right)}={\theta_i^{item}}^{\left(t\right)}-\alpha\left(\frac{\partial L^+}{\partial{\theta_i^{item}}}\right)^{\left(t\right)}$ based on (\ref{eq9})
		  }
		}
	  $t = t + 1$
	}
	\Return{$\theta^{(t)}$}
	\caption{Top-N-Rank}
	\label{alg1}
\end{algorithm}

The gradients for $\Delta_{uji}$ w.r.t. $\theta$ are $\frac{\partial\Delta_{uji}}{\partial\theta_u^{user}}=\left(\theta_j^{item}-\theta_i^{item}\right)$ and $\frac{\partial\Delta_{uji}}{\partial\theta_i^{item}}=-\theta_u^{user}$. $h^\prime$ is the derivative of smooth function which is presented in section~\ref{subsec:smooth_functions}. The pseudo-code for Top-N-Rank (using stochastic gradient descent) is given in Algorithm~\ref{alg1}.

Similar to \cite{shi_climf:_2012}, the computational complexity of Top-N-Rank for one iteration is $O(kn'\tilde{m}^2)$ ($\tilde{m}$ denotes the average number of observed items across all users).

\begin{algorithm}
    \SetKwInOut{Input}{Input}
    \SetKwInOut{Output}{Output}
    \Input{User-item feedback $\mathcal{X}\subseteq\mathcal{U}\times\mathcal{I}\times\mathcal{P}$, the truncate coefficient $N$, smooth function $h$, dimension of latent factors $k$, learning rate $\alpha$, regularization coefficient $\lambda$, batch size $n'$}
    \Output{The learned latent factors $\theta$}
	Initialize $\theta^{\left(0\right)}$ randomly from $U(0,2/ \sqrt[4]{7k})$ and $t=0$ \\
	\While{not converged}{
		$U^{(t)}$ = draw $n'$ samples randomly from $\mathcal{U}$ \\
		\For{$u\in U^{(t)}$}{
		  $\pi$= the descending orders of items indicated by the predicted score $f_u^+$ \\
		  \For{$i={2,\dots,|\pi|}$}{
			$R_{u\pi_i}^+=\sum_{j<i}\left(f_{u\pi_j}-f_{u\pi_i}\right)$ \\
			$\sum_{j\in\mathcal{I}_u^+}{h^\prime\left(\Delta_{uj\pi_i}\right)\frac{\partial\Delta_{uj\pi_i}}{\partial\theta_u^{user}}}=\sum_{j<i}\left(\theta_{\pi_j}^{item}-\theta_{\pi_i}^{item}\right)$ \\
			compute and add the current gradient to $\frac{\partial L^+}{\partial{\theta_u^{user}}}$ based on (\ref{eq9})
		  }
		  Update ${\theta_u^{user}}^{\left(t+1\right)}={\theta_u^{user}}^{\left(t\right)}-\alpha\left(\frac{\partial L^+}{\partial{\theta_u^{user}}}\right)^{\left(t\right)}$ based on (\ref{eq9}) \\
		  $\pi$= the descending orders of items indicated by the predicted score $f_u^+$ \\
		  \For{$i={2,\dots,|\pi|}$}{
			$R_{u\pi_i}^+=\sum_{j<i}\left(f_{u\pi_j}-f_{u\pi_i}\right)$ \\
			$\sum_{j\in\mathcal{I}_u^+}{h^\prime\left(\Delta_{uj\pi_i}\right)\frac{\partial\Delta_{uj\pi_i}}{\partial\theta_{\pi_i}^{item}}}=\left(1-i\right)\theta_u^{user}$ \\
			Update ${\theta_{\pi_i}^{item}}^{\left(t+1\right)}={\theta_{\pi_i}^{item}}^{\left(t\right)}-\alpha\left(\frac{\partial L^+}{\partial{\theta_{\pi_i}^{item}}}\right)^{\left(t\right)}$ based on (\ref{eq9})
		  }
		}
	  $t = t + 1$
	}
	\Return{$\theta^{(t)}$}
	\caption{Top-N-Rank.ReLU}
    \label{alg2}
\end{algorithm}

\subsubsection{Enhanced LTR Algorithm for Large-scale Top-N Recommendation}
\label{sec:Enhance}
Algorithm~\ref{alg1} can be intractable in large-scale systems with massive number of items. The use of ReLU permits a more efficient version of Top-N-Rank (denoted as Top-N-Rank.ReLU) to further reduce the complexity to $O(kn'\tilde{m})$. The pseudo-code for Top-N-Rank.ReLU is given in Algorithm~\ref{alg2}.

For a single user, step 5 and 12 is computed in $O(k\tilde{m}+\tilde{m}\log{\tilde{m}})$. Note that $R_{u \pi_i}^+$ (step 7 and step 14) and $\sum_{j\in\mathcal{I}_u^+}{h^\prime\left(\Delta_{uj\pi_i}\right)\frac{\partial\Delta_{uj\pi_i}}{\partial\theta_u^{user}}}$ (step 8 and step 15) can be calculated in $O\left(1\right)$ and $O\left(k\right)$ respectively through step-by-step accumulation, the complexity of step 6-11 and step 13-17 are $O(k\tilde{m})$. Therefore, the overall computational complexity of Top-$N$-Rank.ReLU for one iteration is $O(n'\tilde{m}(k+\log\tilde{m}))$. In practice, $\log{\widetilde{m}}$ is usually very small (less than 20) even in large-scale systems, Thus, we can expect that $k$ is of the same scale with $\log{\widetilde{m}}$, then the complexity is simplified to $O(kn'\tilde{m})$, making Top-N-Rank.ReLU suitable for large-scale settings with massive item sets.

\section{Experiments and Results}
We report results of two sets of experiments. The first set of experiments compare the performance of Top-N-Rank models using either the sigmoid and the ReLU functions for smoothing with or without the ``top-N truncation''. Our results show that Top-N-Rank.ReLU (using ``top-N truncation'' and the ReLU function, i.e., Algorithm \ref{alg2}) outperforms the other methods on both the benchmark data sets. The second set of experiments compare the performance of Top-N-Rank.ReLU with several state-of-the-art list-wise LTR CF approaches. Our results show that Top-N-Rank algorithms outperform the these methods on both the benchmark data sets.

All of our experiments were performed on an Apache Spark cluster \cite{shoro_big_2015} with four compute nodes (Intel Xeon 2.1 GHz CPU with 20G RAM per node) with the raw data stored on Hadoop Distributed File System (HDFS). The model parameters were tuned to optimize performance on the training data. We describe below the details of the experiments and the results.
\subsection{Experimental Setup}

\subsubsection{Data Sets}

We used two benchmark data sets in our experiments: (i) the Amazon video games data set \cite{mcauley_addressing_2016}, which contains a subset of video games product reviews (ratings, text, etc.) from Amazon. There are 7,077 users, 25,744 items and more than 1 million ratings in this data set. (ii) The MovieLens (20M) data set \cite{harper_movielens_2016}, which contains 138,493 users, 27,278 items and more than 20 millions of ratings. The ratings in both data sets are split to 1-5 stars, with more stars corresponding to higher ratings. We use only the user rating data to conduct the experiments.

\subsubsection{Evaluation Procedure}

We first remove users who rated fewer than 10 items. For the remaining users, we convert the ratings to implicit feedback based on the item ratings provided by each user. That is, for each  $u$, we assign $w_{p_{ui}}=1$ when $f_{ui}\geq4$ and otherwise $w_{p_{ui}}=-1$ \cite{lim_top-n_2015}. We randomly select half of the ratings provided by each user for training, and use the rest for evaluation. On each test run, we average the performance over all of the users. We repeat this process 5 times and report the performance averaged across the 5 independent experiments.

\begin{table*}[tb]
  \centering
  \caption{Comparison of Variants of Top-N-Rank}
  \label{tab2}
  \begin{tabular}{ccccccc}
    \toprule
    Data sets & Algorithms & NDCG@1 & NDCG@3 & NDCG@5 & NDCG@10 & NDCG@20 \\
    \midrule
	 \multirow{5}{*}{Amazon Video Games} & Top-N-Rank.ReLU & \textbf{0.8186} & \textbf{0.8009} & \textbf{0.8079} & \textbf{0.8334} & \textbf{0.8455} \\
	 & non-Top-N.ReLU & 0.8033 & 0.7866 & 0.7976 & 0.8242 & 0.8350 \\
	 & Top-N-Rank.sgm & 0.7956 & 0.7747 & 0.7861 & 0.8167 & 0.8269 \\
	 & non-Top-N.sgm & 0.7871 & 0.7657 & 0.7780 & 0.8109 & 0.8196 \\
	 \midrule
	\multirow{5}{*}{MovieLens} & Top-N-Rank.ReLU & \textbf{0.7811} & \textbf{0.7648} & \textbf{0.7532} & \textbf{0.7469} & \textbf{0.7521} \\
	 & non-Top-N.ReLU & 0.7775 & 0.7593 & 0.7466 & 0.7389 & 0.7430 \\
	 & Top-N-Rank.sgm & 0.7784 & 0.7564 & 0.7415 & 0.7323 & 0.7346 \\
	 & non-Top-N-Rank.sgm & 0.7575 & 0.7315 & 0.7121 & 0.6968 & 0.6954 \\
    \bottomrule
  \end{tabular}
\end{table*}

We measure the performance based only on the rated items as in \cite{lim_top-n_2015}. Because we focus on the placement of the top-rated items in the rank-ordered list, it is natural to use the Normalized Discounted Cumulative Gain (NDCG) \cite{jarvelin_ir_2000} as the performance measure. In this paper, we report the average of NDCG@1 through NCDG@N across all users. 

The definition of NDCG at the top-N positions for a user $u$ is given by:

\begin{equation}
  \label{eq10}
  \mathrm{NDCG}_u\mathrm{@N}=\frac{\mathrm{DCG}_u\mathrm{@N}}{\mathrm{IDCG}_u\mathrm{@N}}
\end{equation}
where DCG$_u$@N is the DCG value for the top-N ranked items as described in (\ref{eq2}). IDCG$_u$@N is the perfect ranking score which is obtained when the ranked list is created by sorting the items  in  descending order of their implicit feedback values (ratings). 

\subsection{Comparison of Variants of Top-N-Rank}
We compare the performance of LTR models trained with the smoothed and regularized wDCG@N objective using either the sigmoid and the ReLU functions for smoothing, with or without the ``top-N truncation'': (i)
\textbf{Top-N-Rank.ReLU}: our proposed Top-N-Rank model trained to optimize wDCG@N smoothed using the ReLU function (Algorithm \ref{alg2}); (ii)
\textbf{non-Top-N.ReLU}: The LTR model trained to optimize wDCG smoothed using the ReLU; (iii)
\textbf{Top-N-Rank.sgm}: our proposed top-N model trained to optimize wDCG@N smoothed using the sigmoid function (Algorithm \ref{alg1}); and (iv) \textbf{non-Top-N.sgm}: The LTR model trained to optimize wDCG smoothed using the sigmoid function.

In these experiments, we set the number of latent factors $k=10$ and the number of items ranked, $N=20$. For the sigmoid function, $C=7$ and for the ReLU function, $b=2/ \sqrt[4]{7k}$ (see section \ref{sec:smooth_func}). The regularization coefficient $\lambda$ is set to $0.1$; and the batch size $n'$ is set to 10\% of the users in the training data. All methods are run until either maximum iteration $maxR=30$ is reached or sum-of-square distance between parameters of two consecutive runs falls below the threshold $\epsilon=0.1$.

The results of our experiments are summarized in Table~\ref{tab2}. Our results clearly show that the Top-N-Rank models with the ``top-N truncation'' term in the objective function consistently and statistically significantly (based on paired Student's $t$-test) outperform the non top-N counterparts. This confirms our intuition that Top-N-Rank models focus on correctly ordering the top-rated items, and hence are resistant to the cumulative effect (often unreliable) of lower-rated items. The results in Table~\ref{tab2} also show that Top-N-Rank.ReLU substantially outperforms Top-N-Rank.sgm. Moreover, the performance of Top-N-Rank.sgm is comparable to that of Non-Top-N.ReLU. We conclude that the ReLU function, with an appropriate choice of $b$ is better able to more accurately rank the top-rated items. The runtime for Top-N-Rank.ReLU is significantly lower than that of Top-N-Rank.sgm (results not shown), proving the appealing efficiency of Top-N-Rank.ReLU.

\begin{table*}[tb]
  \centering
  \caption{Top-N-Rank.ReLU compared with the state-of-the-art list-wise LTR models}
  \label{tab4}
  \begin{tabular}{ccccccc}
    \toprule
    Data sets & Algorithms & NDCG@1 & NDCG@3 & NDCG@5 & NDCG@10 & NDCG@20 \\
    \midrule
	\multirow{7}{*}{Amazon Video Games} & Top-N-Rank.ReLU & \textbf{0.8135} & \textbf{0.7964} & \textbf{0.8043} & \textbf{0.8325} & \textbf{0.8383} \\
          & MF-ADG & 0.7809 & 0.7646 & 0.7762 & 0.8103 & 0.8178 \\
          & CLiMF & 0.7101 & 0.7137 & 0.7388 & 0.7779 & 0.7829 \\
          & xCLiMF & 0.709 & 0.7131 & 0.7381 & 0.7776 & 0.7823 \\
          & ListRank & 0.7045 & 0.7106 & 0.7367 & 0.7761 & 0.7809 \\
          & ListPMF-PL & 0.7043 & 0.7123 & 0.7376 & 0.7762 & 0.78 \\
          \midrule
    \multirow{7}{*}{Movielens} & Top-N-Rank.ReLU & \textbf{0.7827} & \textbf{0.7665} & \textbf{0.7548} & \textbf{0.7483} & \textbf{0.7531} \\
          & MF-ADG & 0.7301 & 0.6993 & 0.6799 & 0.6681 & 0.6698 \\
          & CLiMF & 0.7459 & 0.7187 & 0.7013 & 0.691 & 0.6943 \\
          & xCLiMF & 0.7609 & 0.7406 & 0.7271 & 0.7193 & 0.7236 \\
          & ListRank & 0.7657 & 0.7423 & 0.7284 & 0.7206 & 0.7232 \\
          & ListPMF-PL & 0.6981 & 0.6715 & 0.659 & 0.6568 & 0.6638 \\
    \bottomrule
  \end{tabular}
\end{table*}

\subsection{Top-N-Rank.ReLU  Compared with the State-of-the-Art List-wise LTR Models}
We compare Top-N-Rank.ReLU with several state-of-the-art list-wise LTR CF approaches: (i)
\textbf{MF-ADG}: An algorithm that optimizes the Averaged Discounted Gain (ADG), which is obtained by averaging the DCG across all users \cite{lim_top-n_2015}. Similar to our work, MF-ADG is designed to work with implicit feedback data sets. The sampling parameter $\gamma$ is fixed at 100; (ii)
\textbf{CLiMF}: A MF model that is designed to work with binarized implicit feedback data sets, which optimizes  mean-reciprocal rank (MRR) \cite{shi_climf:_2012}. Instead of directly optimizing MRR, CLiMF learns the latent factors by maximizing the smoothed lower bound of MRR; (iii)
\textbf{xCLiMF}: An extension of CLiMF that optimizes the expected reciprocal rank (ERR), which is designed to work with graded user ratings \cite{shi_xclimf:_2013}; (iv)
\textbf{ListRank}: A MF model that optimizes the cross-entropy between the distribution of the observed and predicted ratings using top-one probability, which is obtained using the softmax function \cite{shi_list-wise_2010}; and (v)
\textbf{ListPMF-PL}: A list-wise probabilistic matrix factorization method that maximizes the log posterior over the predicted rank order with the observed preference order, using the Plackett-Luce model based permutation probability \cite{liu_list-wise_2014}.

The results of our experiments are summarized in Table~\ref{tab4}. Top-N-Rank.ReLU consistently outperforms the baseline models on both the Amazon Video Game and MovieLens data sets, regardless of the length of recommended item lists. Student's $t$ test further demonstrate the significance of our results (not shown). Although Top-N-Rank.ReLU maximize wDCG on the top-20 items, the results show that the model offers better quality of recommendations across the top 1-20 items relative to the baselines. This may be explained in part by the following limitations of the individual methods: CLiMF and xCLiMF are designed to optimize the smoothed reciprocal rank (RR), which does not fully exploit the user ratings, because of its emphasis on optimizing only a few of the relevant items for each user;  MF-ADG maximizes an approximation of ADG, on a small set of sampled data which may limit the quality of the estimates; ListRank and ListPMF-PL are designed for rating data, but assign the same weight to all items with the same rating. Perhaps more importantly, all of the  methods except Top-N-Rank.ReLU attempt to optimize the ranking over the entire set of the user-rated items, as opposed to only the N top-ranked items, which makes them susceptible to the noise in the ratings of low-ranked items.

\section{Summary and Discussion}
In this paper, we proposed Top-N-Rank, a novel family of list-wise Learning-to-Rank models for reliably recommending the N top-ranked items. The proposed models optimize wDCG@N, a  variant of the widely used cumulative discounted gain (DCG) objective function which differs from DCG in two important aspects: (1) It limits the evaluation of DCG only on the top N items in the ranked lists, thereby eliminating the impact of low-ranked items on the learned ranking function;  and  (2) it incorporates weights that allow the model to learn from multiple kinds of implicit user feedback with differing levels of reliability or trustworthiness. Because wDCG@N is non-smooth, we considered two smooth approximations of wDCG@N, using the traditional sigmoid function and the rectified linear unit (ReLU). We proposed a family of learning-to-rank algorithms (Top-N-Rank) that work with any smooth objective function (e.g., smooth approximations of wDCG@N). We designed Top-N-Rank.ReLU, a more efficient version of Top-N-Rank that exploits the properties of ReLU function to reduce the computational complexity of Top-N-Rank from quadratic to linear in the average number of items rated by users. The results of our experiments using two widely used benchmarks, namely, the Amazon Video Games data set and the MovieLens data set demonstrate that: (i) The ``top-N truncation'' of the objective function substantially improves the ranking quality; (ii) using the ReLU for smoothing the wDCG@N objective function yields significant improvement in both ranking quality as well as runtime as compared to using the sigmoid function; and (iii) Top-N-Rank.ReLU substantially outperforms the state-of-the-art list-wise ranking CF methods (MF-ADG, CLiMF, xCLiMF, ListRank, and ListPMF-PL) in terms of ranking quality.

Some promising directions for further research include: (i) Fusing the proposed top-N truncation component and ReLU smoothing function with different list-wise LTR objectives (i.e., MAP, AUC or MRR); (ii) investigation of complex interaction structure of user-item pairs with the help of deep neural nets; (iii) extending the proposed model to tensor factorization or factorization machines to take in multiple types of features.

\section*{Acknowledgment}
Dr. Jinlong Hu and Dr. Shoubin Dong were supported in part by the Scientific Research Joint Funds of Ministry of Education of China and China Mobile [No.~MCM20150512], and the Natural Science Foundation of Guangdong Province of China [No.~2018A030313309]; Junjie Liang was supported in part by a research assistantship funded by the National Science Foundation through the grant [No.~CCF 1518732] to Dr. Vasant G. Honavar. Dr. Vasant Honavar was supported in part by the Edward Frymoyer Endowed Chair in Information Sciences and Technology at Pennsylvania State University, and in part by the Sudha Murty Distinguished Visiting Chair in Neurocomputing and Data Science at the Indian Institute of Science. 

\bibliographystyle{IEEEtran}
\bibliography{b}

\end{document}